\documentclass[prb,aps,floatfix,twocolumn,10pt]{revtex4-2}
\usepackage{graphicx,amsmath,amssymb,color,nicefrac}
\usepackage[utf8]{inputenc}
\usepackage[colorlinks,bookmarks=true,citecolor=blue,linkcolor=red,urlcolor=blue]{hyperref}
\newcommand{\e}{{\rm e}}
\newcommand{\inter}{{\rm inter}}
\newcommand{\intra}{{\rm intra}}
\begin{document}

\title{Eliashberg theory for dynamical screening in bilayer exciton condensation}
\author{G. J. Sreejith$^{1,2}$, Jay D. Sau$^2$ and Sankar Das Sarma$^2$}
\affiliation{$^1$Indian Institute of Science Education and Research, Pune 411008, India }
\affiliation{$^2$Condensed Matter Theory Center and Joint Quantum Institute, Department of Physics, University of Maryland, College Park, Maryland 20742--4111, USA}
\date{\today}
\begin{abstract}
We study the effect of dynamical screening of interactions on the transition temperatures ($T_c$) of exciton condensation in a symmetric bilayer of quadratically dispersing electrons and holes by solving the linearized Eliashberg equations for the anomalous interlayer Green's functions. We find that $T_c$ is finite for the range of density and layer separations studied, decaying exponentially with interlayer separation. $T_c$ is suppressed well below that predicted by a Hartree Fock mean field theory with unscreened Coulomb interaction, but is above the estimates from the statically screened Coulomb interaction. Furthermore, using a diagrammatic framework, we show that the system is always an exciton condensate at zero temperature but $T_c$ is exponentially small for large interlayer separation. 
\end{abstract}
\maketitle
The possibility of Bose-Einstein condensation (BEC) of interlayer excitons formed due to the attractive interaction between electrons and holes in oppositely doped semiconductor bilayers~\cite{keldysh1964, Cloizeux1965, Jerome1967, lozovik1975, Comte1982} has been studied extensively over the past six decades, however, experimental evidence in semiconductor systems outside the quantum Hall regime~\cite{eisenstein2014exciton} has been absent. Recent advances in precision control of particle-hole symmetric electron-hole bilayers have brought us closer to their realization~\cite{Li2017, wang2019, ma2021, chen2022, sun2022, zhang2022, Zijin2023, Davis2023} making conceptual questions of their phase diagram of current relevance.

In addition to the condensate phase, the electron hole bilayer system could exhibit a rich phase diagram containing supersolid, density wave, Wigner crystal, FFLO and Fermi liquid phases as the temperature, interlayer separation ($d$), electron density, screening, interlayer tunneling and layer imbalance are tuned~\cite{Joglekar2006,ZhuBishop2010,parish2011supersolidity}. However the precise zero temperature phase diagram of the symmetric bilayer is not yet settled. Different calculational approaches agree when $d$ is less than the intra layer particle separation -- electrons and holes pair up across the layers to form excitons that undergo a BEC below some finite temperature $T_c$. The attractive interaction weakens rapidly with $d$ and there are two possible scenarios in the opposite limit of large $d$ -- (a) an exciton condensate is always formed albeit at a vanishingly small $T_c$ or (b) the exciton condensate is very unstable and forms an electron-hole plasma even at zero temperature (potentially due to intra-layer screening). Mean field calculations employing different static-screening approximations produce different results.  Unscreened~\cite{Littlewood1996,Zhu1995} as well as static-screening approximations based on normal state correlations show a zero-temperature excitonic condensate phase~\cite{Neilson2014,Kharitonov2008} whereas approximations that incorporate condensate-like correlations in the screening suggest a critical separation beyond which the condensate is absent. Variational and diffusion Monte Carlo studies using ansatz wavefunctions also indicate an electron-hole plasma~\cite{DePalo2002, Maezono2013, Neilson2014,Rios2018, Sara2023}. Similar results are also obtained in graphene-like linear dispersion systems at large $d$ (or large carrier density)~\cite{Hongki2008,Nilsson2021, Perali2013}. 

In the present work, we go beyond the simple Hartree-Fock and static-screening approximations and quantitatively investigate the effect of dynamical screening using the dynamical random phase approximation (RPA) on the $T_c$ by solving the Eliashberg equations~\cite{Marsiglio2020, Parks2018} for the anomalous interlayer Green's functions taking into account the corrected quasiparticle residue, while restricting to the screening calculated from the normal state polarization. Since Eliashberg theory is the most complete theory for calculating $T_c$ in superconductors, with great success in predicting quantitatively accurate $T_c$~\cite{Chubukov2020}, for many superconducting materials, our work is of great significance in understanding electron-hole condensation in 2D bilayer structures. In addition, we show that the diagrams contained in the Eliashberg framework can be extended in the Fermi liquid regime to show that the excitonic instability is robust for arbitrarily weak interlayer interactions.

Our model consists of electrons (represented by $\psi_1$ of charge $-\e$) in one layer $(l=1)$ and holes ($\psi_2$ of charge $\e$) in the other layer $(l=2)$, separated by a distance $d$. In order to address the question of the symmetric bilayer with no interlayer tunneling, we assume identical quadratic dispersons and chemical potentials in the two layers and ignore spin degrees of freedom (changing dispersion does not modify any of our qualitative conclusions). The carriers interact via the attractive (repulsive) Coulomb coupling in the interlayer (intralayer) channel.
The Hamiltonian for the system is given by
\begin{equation}
    H=H_{\rm 0}+ H_{\intra} + H_{\inter}
\end{equation}
where 
\begin{equation}
H_{\rm 0}=\sum_{l=1,2}\int_k (\varepsilon_k-\mu)\psi_{l,\vec k}^{\dagger}\psi_{l,\vec k}
\end{equation}
where the dispersion is given by $\varepsilon_k=\frac{\hbar^2k^2}{2m}$, $\mu=\varepsilon_{K_F}$ is the bare Fermi energy, $K_F$ is the Fermi wavevector and $\int_{{k_1,k_2,\dots}}$ represents $\prod_{i=1,\dots}{d^2k_i}/{{(2\pi)}^2}$.

The intra and interlayer interaction Hamiltonians are 
\begin{align}
    H_{\intra} &= \frac{1}{2} \sum_{l=1,2}  \int_{k,p,q}  V_q \psi_{l,\vec k+\vec q}^\dagger \psi_{l,\vec{p}-\vec{q}}^\dagger \psi_{l,\vec p} \psi_{l,\vec k}\nonumber\\
    H_{\inter} &= -  \int_{k,p,q}  U_q \psi_{1,\vec k+\vec q}^\dagger \psi_{2,\vec{p}-\vec q}^\dagger \psi_{2,\vec p} \psi_{1,\vec k}.
\end{align}
The bare intralayer repulsion ($V_q>0$) and interlayer attraction ($U_q>0$) are 
\begin{align}
    V_q = \frac{\e^{2}}{2\epsilon}\frac{1}{q},\;\;U_q = \frac{\e^{2}}{2\epsilon} \frac{e^{-qd}}{q}
    \label{eq:bareInteractions}
\end{align}
where $\epsilon$ is the average dielectric constant. We assume that both layers have background lattices of neutralizing charges (e.g.\ gates) that cancel the Hartree terms.

A simple fermion renormalization group (RG) analysis~\cite{Shankar1994} at one loop suggests an exciton pairing instability at any $d$ because of the Copper pairing induced by the interlayer electron-hole attraction which cannot be made repulsive by any amount of screening. 
Starting from the Fermi liquid fixed points of the intralayer Hamiltonians, the effective attractive coupling  $U_{\rm qp}>0$ between the quasiparticles of the layers at the wavevector $2K_F$ evolves under RG at one loop as
\begin{center}
    \includegraphics[width=0.9\columnwidth]{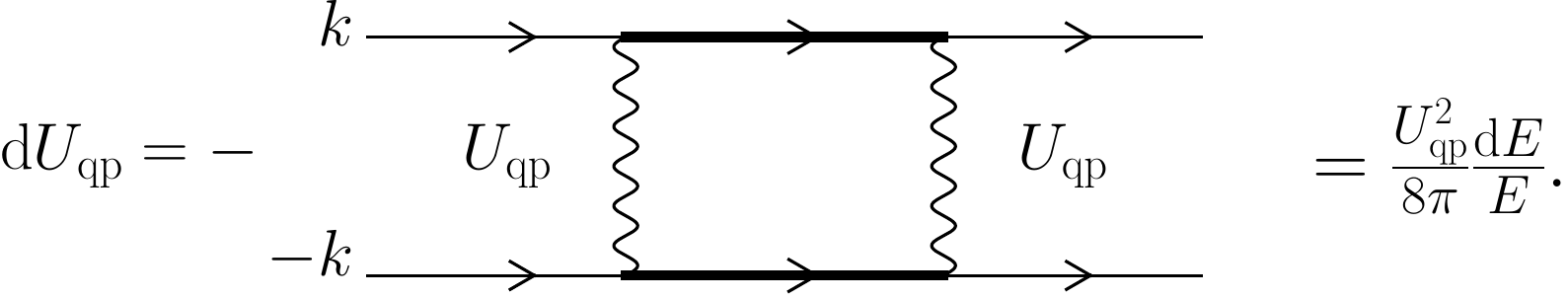}
\end{center}
where the thick lines indicate integration over a thin momentum shell of energy between $E$ and $E+{\rm d}E$ above and below the Fermi surface. For any attractive bare coupling between the quasiparticles of the Fermi liquid $u>0$, $U_{\rm qp}$ diverges as $U_{\rm qp}\sim {(1-u \ln E/8\pi)}^{-1}$. We expect this description to remain valid in the limit of large $d$ where the weak interlayer interactions may not affect the initial RG flow from the microscopic intra layer Hamiltonian to the vicinity of the Fermi liquid fixed points. Consequentially, we expect the exciton instability to survive at large $d$ as long as $u$ is negative.
This expectation can be validated by a more rigorous diagrammatic analysis whose details are presented in the Supplementary material~\cite{SM-Exciton}, which shows that the RG expectation is valid at lowest order in interlayer interaction $U$ and to all orders in intralayer interactions provided the Fermi liquid Green function survives. The only difference is that the parameter $u$ is related to the bare interlayer interaction $U$ through some vertex corrections.
The current work carries out a careful quantitative estimate of the effect of the dynamical screening to calculate the transition temperatures and  establishes that the expectation from RG holds true, even when dynamical screening effects are included.

Within RPA the dynamically screened interlayer and intralayer interactions are given by
\begin{center}
    \includegraphics[width=0.68\columnwidth]{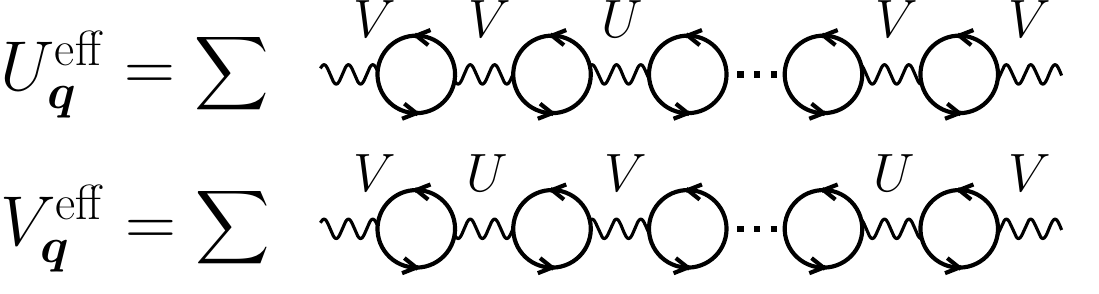}
\end{center}
where the sums are over the number of polarization bubbles ($\Pi$) and also the possible types ($U$ or $V$) of interaction lines between them such that summands of $V_{\boldsymbol{q}}^{\rm eff}$ and $U_{\boldsymbol{q}}^{\rm eff}$ contain even and odd numbers of bare $U_q$ interaction lines respectively. Due to the symmetry of the layers, $\Pi$ is independent of the layer index. The dynamically screened interactions are given by
\begin{align}
    V^{\rm eff}_{\boldsymbol{q}}&=\frac{1}{2}\sum_{\sigma=\pm 1}\frac{V_{q}+\sigma U_{q}}{1+\left(V_{q}+\sigma U_{q}\right)\Pi_{\boldsymbol{q}}}\nonumber\\
    U^{\rm eff}_{\boldsymbol{q}}&=\frac{1}{2}\sum_{\sigma=\pm 1}\sigma \frac{V_{q}+\sigma U_{q}}{1+\left(V_{q}+\sigma U_{q}\right)\Pi_{\boldsymbol{q}}}\label{eq:effectiveInteractions}
\end{align}
We use $\vec p$ for 2D wavevectors, $p$ for their magnitudes and bold font $\boldsymbol{p}$ for the tuple $(\vec p,\imath \omega_p)$ where $\omega_p$ is the Matsubara frequency. The bare polarization function at low temperature is given by (SM~\cite{SM-Exciton}):
\begin{equation}
    \Pi_{\boldsymbol{q}}=\frac{m}{2\pi\hbar^{2}}\left[1-{\rm Re}\sqrt{{\left(1-\imath\frac{\hbar\omega_{q}}{q^{2}\hbar^{2}/2m}\right)}^{2}-\frac{K_{F}^{2}}{q^{2}/4}}\right]
\end{equation}
which is the analytic continuation in the upper-half of complex $\omega$-plane of real-frequency polarization function given in~\cite{Stern1967}.

We define the normal ($\mathcal{G}$) and anomalous ($\mathcal{F}$) imaginary time Green's functions, in the usual manner, as  
\begin{align}
    \mathcal{G}\left(\vec p,\tau\right)&=-\langle T_{\tau} \psi_{l,\vec p}(\tau)\psi_{l,\vec p}^{\dagger}(0)\rangle\nonumber\\
    \mathcal{F}\left(\vec p,\tau\right)&=-\langle T_{\tau} \psi_{1,-\vec p}(\tau)\psi_{2,\vec p}(0) \rangle
\end{align}
The symmetry between the two layers makes the normal Green's function identical in the layers. The normal Matsubara Green's function satisfies the following Dyson equation
\begin{equation}
    \frac{\mathcal{G}_{\boldsymbol{p}}}{\mathcal{G}_{\boldsymbol{p}}^{0}}=1+\frac{1}{\hbar}\Sigma_{\boldsymbol{p}}\mathcal{G}_{\boldsymbol{p}}-\frac{1}{\hbar}W_{\boldsymbol{p}}F_{\boldsymbol{p}}^{\dagger}
\end{equation}
with the normal and anomalous self energies ($\Sigma$ and $W$ respectively) 
\begin{equation}
\Sigma_{\boldsymbol{q}} = \frac{-1}{\hbar\beta }\sum_{\imath\omega_{k}} \int_k V_{\boldsymbol{q}-\boldsymbol{k}}^{{\rm eff}}\,\mathcal{G}_{\boldsymbol{k}},\;
W_{\boldsymbol{q}}=\frac{1}{\hbar\beta }\sum_{\imath\omega_{k}} \int_k  U_{\boldsymbol{q}-\boldsymbol{k}}^{{\rm eff}}\,\mathcal{F}_{\boldsymbol{k}},
\end{equation}
$\beta=1/k_{B}T$, and $\mathcal{G}_{\boldsymbol{k}}$, $\mathcal{F}_{\boldsymbol{k}}$ represent the Matsubara space functions obtained by suitable Fourier transforms of the imaginary time functions $\mathcal{G}(\vec{k},\tau)$, $\mathcal{F}(\vec{k},\tau)$. The exciton condensation is defined by a non-zero anomalous self-energy which is related to the condensate order parameter as $\Delta_{\boldsymbol{q}}\sim W_{\boldsymbol{q}} /Z_{\boldsymbol{q}}$.

Using the Dyson equation, these can be expressed as a set of coupled Eliashberg equations incorporating dynamical screening in the exciton condensate formation. It is convenient to express the normal self-energy in terms of parts that are even and odd in frequency, parametrized by $S$ and $Z$ respectively, as $\Sigma_{\boldsymbol{q}} = S_{\boldsymbol{q}} + \imath\omega_q (1-Z_{\boldsymbol{q}})$. The Eliashberg equations can then be expressed as 
\begin{align}
    Z_{\boldsymbol{q}} &= 1 - \frac{1}{\hbar\beta}\sum_{\imath\omega_{k}}\int_k 
    V_{\boldsymbol{q-k}}^{{\rm{eff}}}\frac{\imath \omega_{k}}{\imath \omega_{q}}\frac{Z_{\boldsymbol{k}}}{\Lambda_{\boldsymbol{k}}}\nonumber\\
    S_{\boldsymbol{q}} &= \frac{1}{\hbar\beta}\sum_{\imath\omega_{k}}\int_k 
    V_{\boldsymbol{q-k}}^{{\rm eff}} \frac{\frac{1}{\hbar}E_{\boldsymbol{k}}}{\Lambda_{\boldsymbol{k}}}\label{eq:EliashbergEquations}\\
    W_{\boldsymbol{q}} &= \frac{1}{\hbar\beta}\sum_{\imath\omega_{k}}\int_k 
    U_{\boldsymbol{q-k}}^{{\rm eff}}\frac{W_{\boldsymbol{k}}}{\Lambda_{\boldsymbol{k}}}\nonumber
\end{align}
where $\Lambda_{\boldsymbol{k}}={{\left[\omega_{k}Z_{\boldsymbol{k}}\right]}^{2}+\frac{1}{\hbar^{2}}\left(E_{\boldsymbol{k}}^{2}+W_{\boldsymbol{k}}^{2}\right)}$ and $E_{\boldsymbol{k}}=\epsilon_k + S_{\boldsymbol{k}}-\mu$.

\begin{figure}
    \includegraphics[width=\columnwidth]{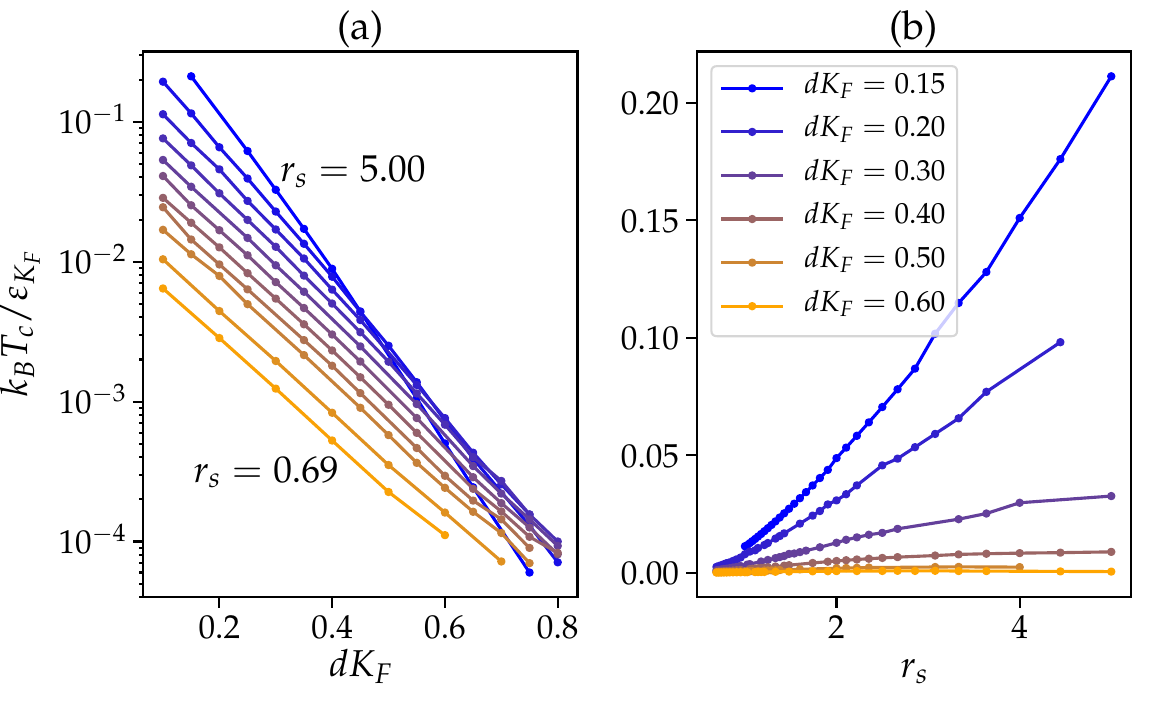}
    \includegraphics[width=\columnwidth]{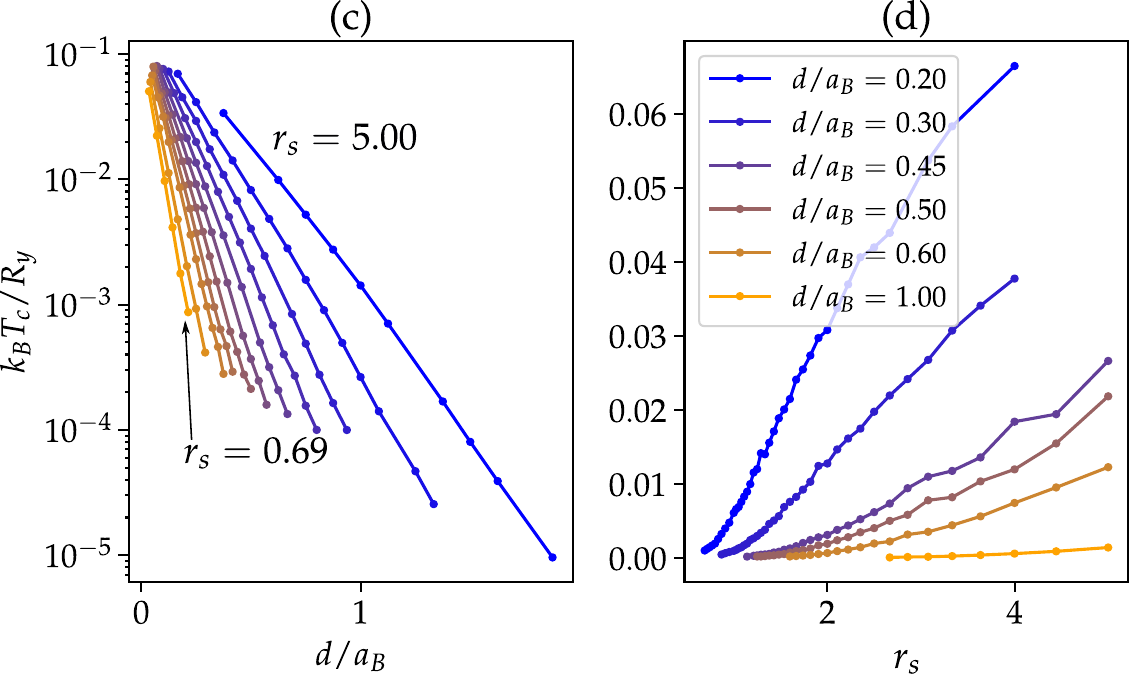}
    \caption{$T_c$  from the screened HF mean field equations (a)  $T_c/T_F$ as a function of the dimensionless interlayer separation $dK_F$ for different values of $r_s$. $T_c$ decreases rapidly but remains finite even for the largest values of $d$. (b) $T_c/T_F$ for fixed values of $dK_F$ increases with $r_s$ except at very large $r_s$ and $dK_F$. (c,d) Similar to (a,b) but in atomic units.\label{fig:screened}}
\end{figure}
The anomalous self-energy $W_{\boldsymbol{q}}$ is zero at high temperatures and becomes finite as the temperature is lowered if there is a transition into an excitonic condensate phase. We estimate the transition temperature $T_c$ by numerically solving a linearized (in $W$) form of the above equations which is valid near $T_c$.
We assume full rotational invariance and consider only $s$-wave solutions, as a result of which $W$, $S$ and $Z$ depend only on $(q,\imath\omega_q)$. 

\textit{Static screening}: Before considering the solution to the Eliashberg equations, we consider a static screening approximation \textendash~we ignore corrections to $Z$, setting it to the bare value $1$, and ignore frequency dependences of $S$, $W$,  $V^{\rm eff}$ and $U^{\rm eff}$. After performing the Matsubara summations in Eq.~\eqref{eq:EliashbergEquations}, we get the following mean field equations
\begin{align}
    \Sigma_{q}&= \frac{1}{2} \int_q V_{q}^{0} \left[\frac{E_q}{\xi_q} \tanh \frac{\beta \xi_{q}}{2}-1 \right]\nonumber\\
    W_{q}&= \frac{1}{2} \int_q U_{q}^{0} \left[\frac{E_q}{\xi_q} \tanh \frac{\beta \xi_{q}}{2} \right]\label{eq:HFMeanField}
\end{align}
where $E_q = \varepsilon_q+\Sigma_q-\mu$ and $\xi = {(E_q^2 + W_q^2)}^{\frac{1}{2}}$. The statically screened interactions $V_{q}^{0}$ and $U_q^{0}$ are the zero frequency limits of the effective interactions shown in Eq.~\eqref{eq:effectiveInteractions}. Figure~\ref{fig:screened} shows the transition temperatures obtained by solving the equations numerically, presented as a function of the Wigner Seitz interaction radius $r_s={2}/{a_{\rm B} K_F}$ (i.e.\ the dimensionless intralayer particle separation), where $a_{\rm B}$ is the Bohr radius. We find that the transition temperatures $T_c$ in Fermi units (Fermi temperature $T_F=\varepsilon_{K_F}/k_B$ and $1/K_F$ as units of temperature and length respectively) decays exponentially  as a function of $d$.
\begin{figure}
    \includegraphics[width=\columnwidth]{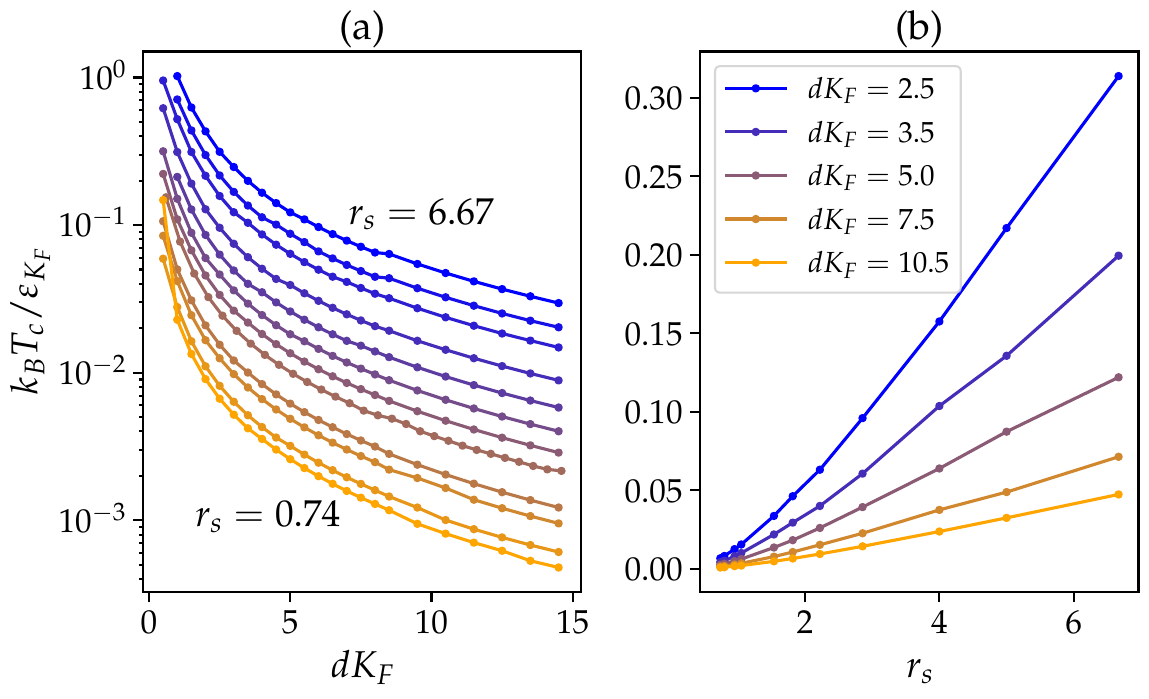}
    \includegraphics[width=\columnwidth]{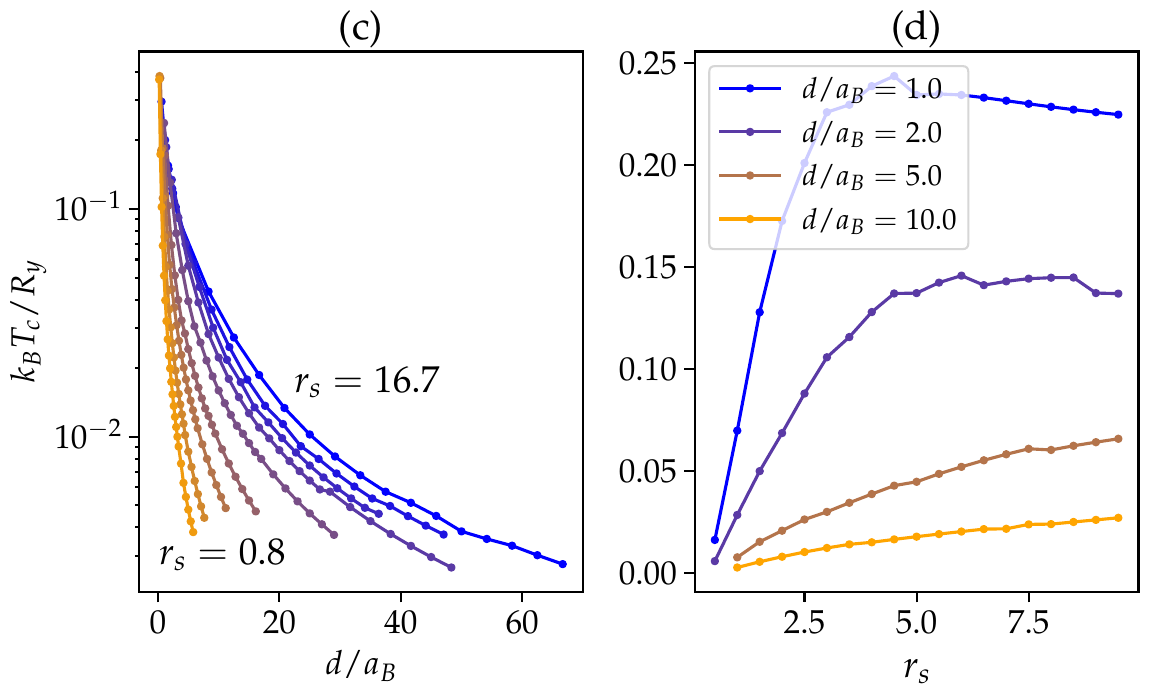}
\caption{$T_c$  from the unscreened HF mean field equations. (a)  $T_c/T_F$ as a function of $dK_F$ for different values of $r_s$. (b) $T_c/T_F$ for fixed separations $dK_F$ increases with $r_s$. (c,d) Similar to (a,b) but with axes in atomic units.\label{fig:unscreened}}
\end{figure}

Equations~\eqref{eq:HFMeanField} are the Hartree-Fock (HF) mean field equations analysed in Ref.~\cite{Zhu1995,Littlewood1996,zhu2023}, except for the use of unscreened interactions. An unscreened approximation can be obtained by replacing the interactions $V^0$and $U^0$ with the bare Coulomb interactions (Eq.~\eqref{eq:bareInteractions}). Figure~\ref{fig:unscreened} shows the $T_c$ obtained by solving the unscreened mean field equations. Panel (a) shows $T_c/T_F$ as a function of $dK_F$. Panel (c) shows the similar data but with the $T_c$ and $d$ in atomic units ($R_y/k_B=\frac{1}{k_B}\frac{\e^{2}}{8\pi\epsilon a_{B}}$ and $a_B$ as units for temperature and length).

In both static approximations (unscreened and statically screened HF), we find that $T_c$ decays rapidly with increasing $d$ approximately exponentially when plotted in Fermi units. 
The statically screened mean field equations predict a very small $T_c$ which is consistent with previous observations in multicomponent systems~\cite{Abergel2012,Kharitonov2008}. The $T_c$ increases with $r_s$ but tends to saturate at very large $r_s$~\cite{Zhu1995}. 
\begin{figure*}[!]
    \includegraphics[width=\textwidth]{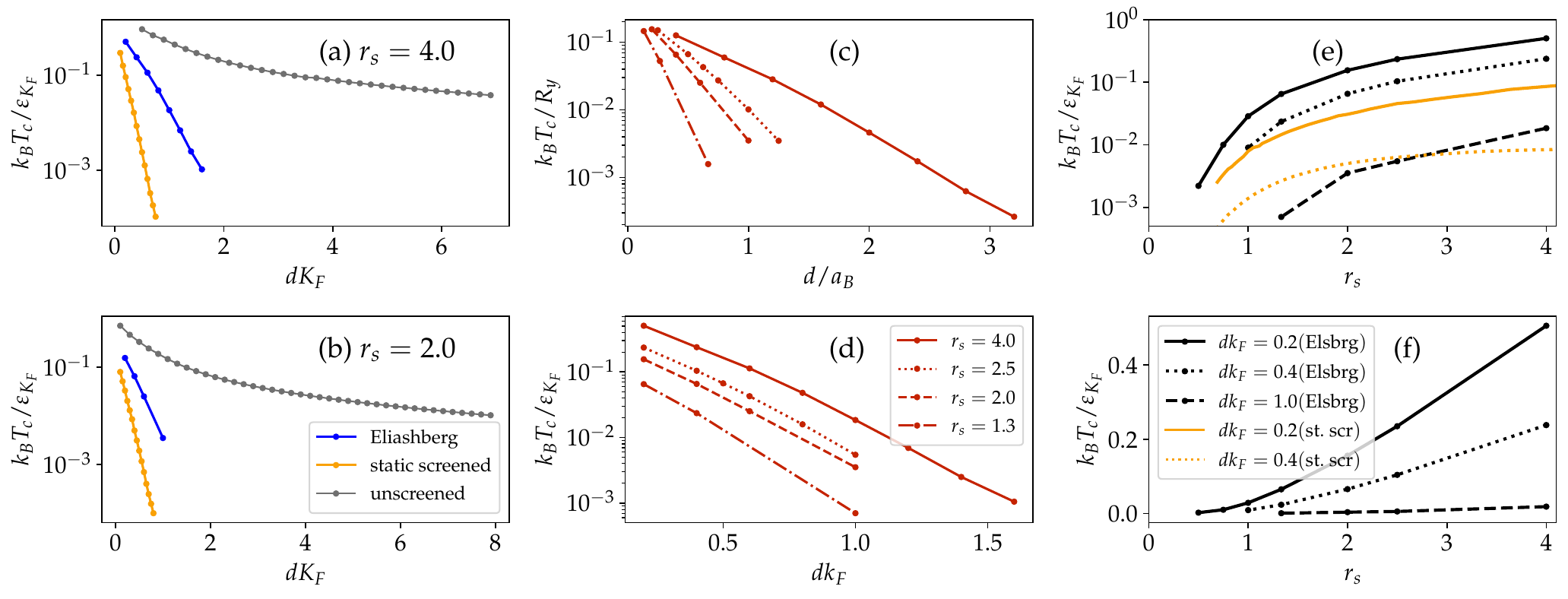}
    \caption{$T_c$ estimates from Eliashberg equations (a-b)~Comparison of the $T_c$ estimated from the Eliashberg equations (blue) and the static screening approximations. (c,d)~$T_c$ for different $r_s$ as a function of $d$ plotted in Fermi and atomic units. (e,f)~$T_c$ as a function of $r_s$ for fixed values of $d$. Orange lines in (e) show the results for the statically screened approximations for the corresponding values of $d$ and $r_s$. Plots in a column share the same legend.\label{fig:eliashberg}}
\end{figure*}

\textit{Dynamical Screening}: To obtain $T_c$ we numerically estimate the magnitude of the eigenvalue that governs the evolution of $W_{\boldsymbol{q}}$ under iterations of the linearized self-consistent Eliashberg equations treated as a recursive relation. The eigenvalue decreases with $T$ and is $1$ at $T_c$. The linearized (in $W$) equations for $S$ and $Z$ are independent of $W$ and can be solved first using iterations, these are found to converge rapidly~\cite{Grankin2023} and these results can be used in the equations for $W$ whose leading eigenvalue can be determined by power iteration method.

Following Ref.~\cite{Rietschel1983}, we find it convenient to decompose the even part of the self energy $S$ into frequency dependent and independent components:
\begin{equation}
    S_{\boldsymbol{q}}=S^{\omega}_{\boldsymbol{q}} + S^{0}_{{q}}
\end{equation}
where the frequency dependent component
\begin{align}
    S^{\omega}_{\boldsymbol{q}} &= \frac{1}{\hbar\beta} \sum_{\imath \omega_k} \int_k (V_{\boldsymbol{k-q}}^{\rm eff}-V_{|\vec{k}-\vec{q}|}) \frac{E_{\boldsymbol{k}}/\hbar}{\Lambda_{\boldsymbol{k}}}
\end{align}
is found to vanish at large wavevectors and frequencies~\cite{Rietschel1983} and therefore has a finite support in the $q-\omega$ plane (the cutoffs for this are set at $q=40K_F$ and $\hbar\omega\sim 50\varepsilon_{K_F}$ in the numerics). Same cutoffs (to reach saturation) are used for $Z_{\boldsymbol{q}}$ which asymptotes to $1$ at large frequencies and wavevectors. Matsubara sums are performed, in all cases at least upto $\hbar \omega_k=135\varepsilon_{K_F}$ even at low temperatures.

The frequency independent component of $S$ can be written as 
\begin{equation}
    S_{{q}}^{0} = \frac{1}{\hbar\beta}  \int_k V_{|\vec{k}-\vec{q}|} \sum_{\imath \omega_k} \left ( \frac{E_{\boldsymbol{k}}/\hbar}{\Lambda_{\boldsymbol{k}}} - \frac{1}{2}\right).
    {\label{eq:exchangeSelfEnergyEven0}}
\end{equation}
The chemical potential $\mu$ is tuned self-consistently to satisfy $E_{K_F}(\omega=0)=0$ keeping the Fermi wavevector fixed. $S^0$ can be efficiently estimated to high accuracy (See SM~\cite{SM-Exciton}).
The gap $W_{\boldsymbol{q}}$ vanishes at large wavevector and saturates to a $q$-dependent constant $W^\infty_{q}$ at large frequency, for which we set a cutoff (to reach saturation) of $\hbar \omega\sim 50\varepsilon_{K_F}$. Representative plots of $S,W$ and $Z$ are shown in the Supplemental materials~\cite{SM-Exciton}.

Figure~\ref{fig:eliashberg} shows the $T_c$ estimated from solving the Eliashberg equations. 
$T_c/T_F$ decreases exponentially but remains finite for all $r_s$ and $d$. Dynamical screening suppresses the $T_c/T_F$ to a value smaller than the estimates from unscreened approximations. Approximate treatment of dynamical screening effects~\cite{Sodemann2012} in graphene have indicated a first order transition as a function of distance. Within the range of distances where we could reliably perform our calculations we do not find such a transition. $T_c$ at fixed $r_s$ exponentially decays with $d$ till the largest value that we could study. At fixed $d$, and decreasing $r_s$, $T_c$ again remains finite up to the smallest $r_s$ we could access.

Eliashberg theory is known to be insufficient when the frequency scales of the attractive interaction (along the real-frequency axis) are comparable to the Fermi energy~\cite{Marsiglio2020}. In our case the dominant frequency content of $U^{\rm eff}$ correspond to the plasmons of the 2D system which have small frequencies on account of its $\hbar \omega/E_F\sim \sqrt{r_s q}$ dispersion. This is especially true at small $r_s$ and large $d$ where the plasmon mode diffuses into the particle-hole continuum as the frequency is increased (See SM~\cite{SM-Exciton} where we show the spectral function of $U^{\rm eff}$ averaged over the Fermi surface).

Our main findings are (1) $T_c$ is finite always, but exponentially small for large $d$; (2) $T_c$ is lower than that obtained from the unscreened HF theory; (3) interlayer coherent exciton condensate exists for all parameters at $T=0$.  The reason for the bilayer to be always interlayer coherent at $T=0$ is that the interlayer interaction is always attractive, and this implies that there is no repulsion-induced $\mu^*$ effect in $T_c$ as in metallic superconductors~\cite{Allen1975}-- all that intralayer interactions can do is to suppress the effective interlayer attraction, but can never make it vanish, implying that $T_c$ is always finite albeit very small for large $d$.
This expectation is argued to be correct beyond Eliashberg theory (i.e.\ including arbitrary intralayer vertex corrections) where a bound on $T_c$ that is comparable to the static screening limit is obtained for the general Bethe-Salpeter equation.

\acknowledgments{G.J.S thanks Yang-Zhi Chou, Andrey Grankin, Darshan Joshi, R Sensarma, A Balatsky and F Marsiglio for useful inputs. 
This work is supported by the Laboratory for Physical Sciences.  One of the authors (G.J.S) acknowledges partial JQI support for a sabbatical leave.
The authors also thank National Supercomputing Mission (NSM), India for providing the computing resources of `PARAM Brahma' at IISER Pune, which is implemented by C-DAC and supported by the Ministry of Electronics and Information echnology (MeitY) and Department of Science and Technology (DST), Government of India.}

\bibliography{exciton.bib}

\end{document}


\title{Supplemental material for Eliashberg theory for dynamical screening in bilayer exciton condensation}
\author{G. J. Sreejith$^{1,2}$, Jay D. Sau$^2$ and Sankar Das Sarma$^2$}
\affiliation{$^1$Indian Institute of Science Education and Research, Pune, India 411008}
\affiliation{$^2$Condensed Matter Theory Center and Joint Quantum Institute, Department of Physics, University of Maryland, College Park, Maryland 20742--4111, USA}
\date{\today}
\maketitle

\section{Introduction}
\subsection{Diagrams for exciton instability at lowest order in interlayer interaction:}

\begin{figure*}[t]
\includegraphics[width=0.7\textwidth]{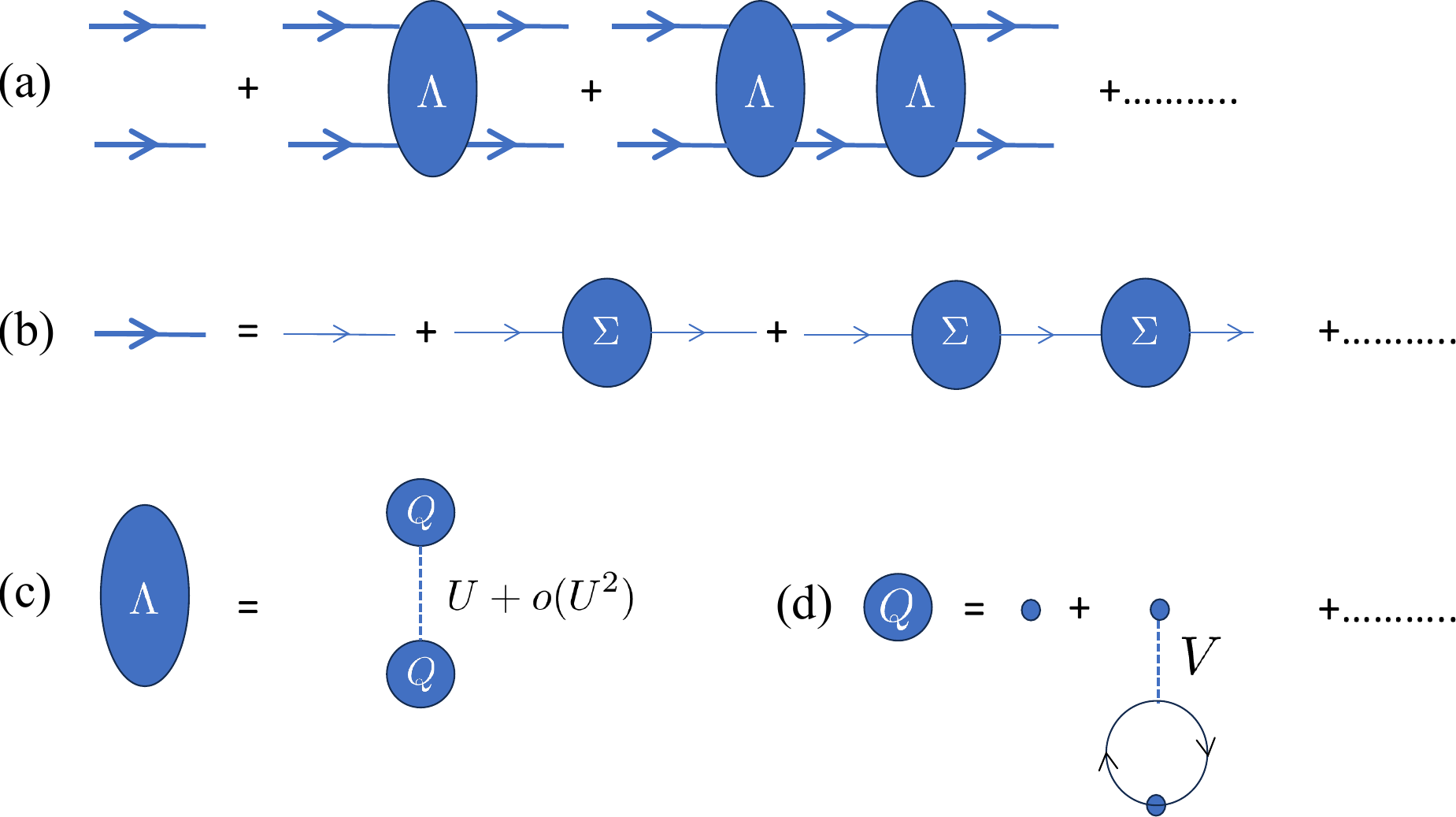}
\caption{\label{fig:diagrams} Organization of diagrams that leads to perturbative i.e. small intra-layer limit of gap equation.   (a) Bethe-Salpeter diagram for electron-hole propagator $G_2^c$ in terms of the effective interaction $\Lambda$. (b) Diagram for the single-layer fermion Green function $G_1$ in terms of the self-energy $\Sigma$ that contains only 
intra-layer interaction $V$. (c) Expansion for the effective interaction $\Lambda$ in terms of inter-layer interaction $U$. (d) Expansion of ``vertex corrections'' $Q$ in terms of intra-layer interaction $V$.}   
\end{figure*}

Exciton condensation is characterized by a spontaneous breaking of the layer $U(1)$ symmetry in a bilayer system and arises from a non-zero expectation value of $\expect{\phi_{\vec{k},i\omega}}=\expect{\psi_1(\vec{k},i\omega )\psi_2(-\vec{k},-i\omega)}$. Such a spontaneous symmetry breaking can be described in terms of the Ginzburg-Landau functional $\Gamma[\phi]$ through the condition $\Gamma^{(2)}[\phi]=\delta^2 \Gamma[\phi]/\delta\phi^2<0$. In quantum field theory, this Ginzburg-Landau functional is called the effective action and $\Gamma^{(2)}[\phi]=G^{c}_2[\phi]^{-1}$ is the inverse of the propagator of $\phi$. Since $\phi$ is a pair of fermions, the connected Green function $G_2^c$ for $\phi_{k,\omega}$ is written in terms of a Bethe-Salpeter equation~\cite{Strinati1988} shown in Fig.~\ref{fig:diagrams}(a). The introduction of the effective interaction $\Lambda$ allows us to write the spontaneous symmetry breaking condition as 
\begin{align}\label{eq:inst1}
&\Gamma^{(2)}[\phi]\varpi=(G_2^c)^{-1}\varpi=[(G_2^{c,0})^{-1}-\Lambda]\varpi=0,
\end{align}
for some choice of the vector $\varpi$ where $G_2^{c,0}=G_{1}^{(1)}G_{1}^{(2)}$ is the electron-hole pair propagator when interlayer interaction $U=0$. Note the above equation is a formal matrix vector equation with various indices and sums suppressed for simplicity. These will be elaborated in following equations.  Here $G_1^{j=1,2}$ are the single particle propagators in each layer $j=1,2$ whose diagrammatic expansion is shown in Fig.~\ref{fig:diagrams}(b). The diagrams for the propagator contain the one particle irreducible single-particle self-energy $\Sigma$.  The effective interaction $\Lambda=\delta\Sigma/\delta G_1$ in the Bethe-Salpeter equation can be calculated diagrammatically by removing one internal fermion leg from all diagrams in the self-energy $\Sigma$~\cite{Strinati1988}.
The diagrams for the  effective interaction $\Lambda$ between fermions of different layers, to lowest order in the interlayer interaction $U$ are shown in Fig.~\ref{fig:diagrams}(c).  Combining all the diagrams in Fig.~\ref{fig:diagrams}(a-c)
and using the fact that the fermion legs on the top and bottom row are electron-hole pairs from different layers  with vanishing total energy and momentum, Eq.~\ref{eq:inst1} simplifies to 
\begin{align}
G_1(\vec k,i\omega)^{-1}G_1(-\vec k,-i\omega)^{-1}\varpi_{\vec k,\omega}=\sum_{\vec k'\omega'}Q(\vec k,\vec k';i\omega,i\omega')Q(-\vec k,-\vec k';-i\omega,-i\omega')U(\vec k-\vec k')\varpi_{\vec k'\omega'},
\end{align}
Assuming that (a) the electron and hole layers are exactly symmetric both in terms of dispersion and interactions, (b) the dispersion and interaction are even functions $\epsilon_{\vec k}=\epsilon_{-\vec k}$, $V(\vec k)=V(-\vec k)>0$ and $U(\vec k)=U(-\vec k)>0$ we can rewrite the above equation in a form closer to the BCS gap equation.
\begin{align}
&\Delta_{\vec k,\omega}=\sum_{\vec k',\omega'}|Q(\vec k,\vec k';\imath\omega,\imath\omega')|^2 |G_1(\vec k',\imath\omega')|^2U(\vec k-\vec k')\Delta_{\vec k,'\omega'},\label{eq:gap}
\end{align}
where the gap function $\Delta_{\vec k,\omega}=\varpi_{\vec k,\omega}|G_1(\vec k,\imath\omega)|^{-2}$. This equation is essentially identical to the Eliashberg equation for the gap i.e. the equation for $W\sim \Delta$ in Eq.~10 in the main text, if we choose $U^{\rm eff}=|Q|^2U$.

The factor $Q(\vec k, \vec k';i\omega, i\omega')$ is a vertex correction which contains diagrams such as the ones shown in Fig.~\ref{fig:diagrams}(d). In fact, the diagrams for $Q$ can be figured out from the rules by considering all diagrams for $\Sigma$ for layer $1$ that contain only one interlayer interaction $U$ and then removing one propagator for layer 2, according to the definition of the vertex function $\Lambda$. From these rules it is clear that $Q$ in addition to other contributions contains the dynamical screening that is used in the Eliasberg equations solved in the main text. This is 
the main advantage of this diagram approach over the RG approach discussed in the main text. The interaction $U^{\rm eff}$ would replace the parameter $u$ in the RG framework.

\subsection{Lower bound on \texorpdfstring{$T_c$}{Lg} for excitonic instability}
As is clear from the main-text, solving the above gap equation for $T_c$ is quite non-trivial. On the other hand, it is much 
easier to establish lower bounds on $T_c$. To do this, we first separate $U(\vec k-\vec k')=U_0 \Phi(\vec k-\vec k')$, where $U_0$ is an overall magnitude. The idea is that it is easier to compute $U_0$ as a function of $T$ then the reverse.

This allows us to write  the gap equation (Eq.~\ref{eq:gap}) as an eigenvalue problem
\begin{align}
U_0^{-1}\Delta_{\vec k,\imath\omega}=\sum_{k',\omega'}\Phi(\vec k-\vec k')|Q(\vec k,\vec k';\imath \omega,\imath\omega')|^2|G_1(\vec k',\imath\omega')|^2 \Delta_{\vec k,\imath\omega'},
\end{align}
where the largest eigenvalue $U_0^{-1}$ solves the inverse problem of determining the interaction for which the temperature of the calculation is the transition temperature i.e. $T_c(U_0)=T$. However, this is not a symmetric eigenvalue problem, which are the class of problems that can be diagonalized in general. To resolve this, let us first note that the vertex function $Q(\vec k,\vec k';i\omega,i\omega')$ can be checked to be symmetric based on the contributing diagrams in Fig.~\ref{fig:diagrams}(d). Next we change variables to $\zeta_{\vec k,i\omega}=|G_1(\vec k,i\omega)|\Delta_{\vec k,i\omega}$, so that 
the eigenvalue problem is now symmetric:
 \begin{align}
U_0^{-1}\zeta_{\vec k,\imath\omega}=\sum_{k',\omega'}\Phi(\vec k-\vec k')|Q(\vec k,\vec k';\imath \omega,\imath \omega')|^2 |G_1(\vec k',i\omega') G_1(\vec k,\imath\omega)| \zeta_{\vec k',\imath\omega'}.
\end{align}
Additionally, since the relevant matrix is a positive symmetric matrix, the eigenvector corresponding to the largest eigenvalue $U_0^{-1}$ is positive (i.e. $\zeta_{\vec k,i\omega}>0$) everywhere.  While the largest eigenvalue is non-trivial to calculate analytically, it can be bounded below by the expectation value 
\begin{align}\label{eq:variation}
E[\{\zeta \}]=\sum_{k\omega,k'\omega'}\Phi(\vec k-\vec k')|Q(\vec k,\vec k';\imath \omega,\imath \omega')|^2
|G_1(\vec k',\imath\omega')| |G_1(\vec k,\imath\omega)|\zeta_{\vec k',\imath\omega'}\zeta_{\vec k,\imath\omega}.
\end{align}
 for any choice of normalized positive vectors $\zeta$.

The remarkable property of  the BCS problem is that the largest eigenvalue of the matrix in the case of a Fermi liquid has a log divergence as $T\rightarrow 0$, where $T$ determines the Matsubara frequencies $\omega=(2 n+1)\pi T$. To show this 
bound we will restrict our ansatz  $\zeta_{\vec k,i\omega}>0$ for computing the bound to be non-zero only in a thin shell of energy $|k-k_F|<\Xi$, $|k'-k_F|<\Xi$ and $|\omega|<\Xi$ where the width is chosen so that $T\ll \Xi\ll E_F$. Note 
that this is not an approximation since Eq.~\ref{eq:variation} still provides a bound on $U_0$ for any normalizable choice for  $\zeta_{\vec k,i\omega}$. Within this limit we can assume that the Green-function is approximated by Fermi-liquid theory so that $|G_1(\vec k,i\omega)|\simeq \frac{Z( k)}{\sqrt{\omega^2+\tilde{\epsilon}^2_{\vec k}}}$, where $\tilde{\epsilon}(\vec k)$ 
is the quasiparticle dispersion and $Z( k)$ the quasiparticle weight. We will further assume that the frequency dependence 
of the vertex correction $|Q(\vec k,\vec k';\imath \omega,\imath \omega')|$ can be ignored since $|\omega|\ll E_F$. Strictly speaking this assumption can be violated in two dimensions over a small part of the Fermi surface because of the gapless plasmon.

The BCS lower bound is obtained by choosing $\zeta_{\vec k,\imath\omega}=R/\sqrt{(\omega^2+\tilde{\epsilon}_k^2)}$ where it is non-zero as discussed in the previous paragraph. Here $R$ is a normalization constant. We first perform the angular integrals of $\Phi$ and $Q$. We assume that these integrals and $Z$ are constants in the support of  $\zeta$ and the Fermi surface is rotationally symmetric. We define this constant to be 
\begin{align}
F(\epsilon=v_F(k-k_F),\epsilon'=v_F(k'-k_F))=Z^{1/2}(k)Z^{1/2}(k')\int d\theta d\theta' \Phi(\vec k-\vec k')|Q(\vec k,\vec k';\imath\omega,\imath\omega')|^2\approx F, 
\end{align}
over the range $|\epsilon|,|\epsilon'|<\Xi$ and $v_F$ is the Fermi velocity of the interacting Fermi liquid in each 
layer.
The eigenvalue bound Eq.~\ref{eq:variation} can be written in terms of $F$ (after performing the Matsubara sums) as 
\begin{align} \label{eq:EF}
E[\{\zeta \}]=N(0)^2 R^2\int d\epsilon d\epsilon'\frac{F(\epsilon,\epsilon')\tanh{(\beta \epsilon/2)}\tanh{(\beta\epsilon'/2)}}{\epsilon\epsilon'}
\simeq N(0)^2 R^2 F \log[\Xi/T]^2,
\end{align}
where $N(0)$ is a density of states factor.
The normalization factor  $R^2$ is determined by the equation 
\begin{align}
N(0)R^2\int d\epsilon \sum_\omega (\omega^2+\epsilon^2)^{-1}=R^2 N(0)\int d\epsilon \frac{\tanh{(\beta\epsilon/2)}}{\epsilon}\sim R^2 N(0) \log{(\Xi/T)}=1.
\end{align}
Substituting the factor $R^2$ cancels one factor of logs in Eq.~\ref{eq:EF} so that the eigenvalue bound becomes 
\begin{align}
U_0^{-1}>E[\{\zeta\}]\simeq  N(0) F \log[\Xi/T].
\end{align}
This leads to the bound on $T_c$ 
\begin{align}\label{eq:Tcbound}
&T_c\gtrsim \Lambda e^{-1/(N(0)F |U_0|)},
\end{align}
which is of the BCS form, assuming an attractive interaction (i.e. $U_0>0$ in the chosen convention).
This shows that the log excitonic instability is generic for the Fermi liquid. Note that this formula depends on a 
somewhat arbitrary scale $\Xi$. If $\Xi$ is chosen to be too small the $T_c$ would be an underestimate. In our 
case $\Xi$ should be of order Fermi energy, which is the scale over which the frequency dependence of the Fermi function should be significant. Since we have chosen the frequency cutoff of our ansatz to be $\Xi\ll E_F$, the effective interaction $|U_0| F$ corresponds to the static screened results in the main text.  However, chosing a larger $\Xi$ would make the frequency dependence siginificant making the bound less reliable. A more correct estimate of $T_c$ requires a solution of the gap equation as presented for 
the RPA case in the main text. The result Eq.~\ref{eq:Tcbound} serves the main purpose of providing a bound that shows that the exciton instability survives to arbitrarily small $U_0$.

\section{Polarization\label{app:polarization}}
In this section, we present the estimation of the bare polarization $\Pi_{\boldsymbol{q}}$ for the 2D electron gas that appears in Eq~5 of the main text. $\Pi_{\boldsymbol{q}}$ is given by
\begin{equation}
    \Pi_{\boldsymbol{q}}=-\frac{1}{\hbar^{2}\beta}\sum_{\imath\omega_{k}}\int_k\mathcal{G}_{k}^{0}\mathcal{G}_{|\vec{k}+\vec{q}|}^{0}=\\-\frac{1}{\hbar\beta}\sum_{\imath \hbar\omega_{k}}\int_k\frac{1}{(\imath \hbar\omega_{k}-\varepsilon_{k}+\mu)(\imath \hbar\omega_{k}+\imath\omega_{q}-\varepsilon_{|\vec{k}+\vec{q}|}+\mu)}.
\end{equation}
The Matsubara sum in the above equation can be performed so that $\Pi_q$ is written as  
\begin{equation}
    \Pi_{\boldsymbol{q}}=-\int_k \frac{n_{\varepsilon_{k}-\mu}^{F}}{\imath\hbar\omega_{q}+\varepsilon_{k}-\varepsilon_{|\vec{k}+\vec{q}|}}-\frac{n_{\varepsilon_{k}-\mu}^{F}}{\imath\hbar\omega_{q}+\varepsilon_{|\vec{k}-\vec{q}|}-\varepsilon_{k}}
\end{equation}
where $n^F_{\varepsilon}={(1+e^{\beta \varepsilon})}^{-1}$ is the Fermi distribution function. The integral over the angular part of $\vec{k}$ can be performed using 
\begin{equation}
    \int_{0}^{2\pi}\frac{d\theta}{2\pi}\frac{1}{\cos\theta+w}=\frac{{\rm sign}\left({\rm Re} w\right)}{\sqrt{w^{2}-1}}
\end{equation}
so that we get the result
\begin{equation}
    \Pi_{\boldsymbol{q}} = \int_0^\infty \frac{kdk}{2\pi}{\rm Re}\frac{2n^F_{\varepsilon_k-\mu}}{\sqrt{{\left(\frac{q^{2}\hbar^{2}}{2m}-\imath\hbar\omega_{q}\right)}^{2}-\frac{\hbar^{4}q^{2}k^{2}}{m^{2}}}}.
\end{equation}
In the low temperature limit, we can approximate, $n^F_{x}$ with $\theta(-x)$ and get the following expression for the polarization diagram (with the convention that the square root has a positive real part).
\begin{equation}
    \Pi_{\boldsymbol{q}} = \frac{m}{2\pi\hbar^{2}}\left[1-{\rm Re}\sqrt{{\left(1-\imath\frac{\hbar\omega_{q}}{q^{2}\hbar^{2}/2m}\right)}^{2}-\frac{K_{F}^{2}}{q^{2}/4}}\right]
\end{equation}

The zero temperature polarization in the real frequency space can be calculated in a similar manner and gives
\begin{equation}
\Pi^R(q,\omega)=\frac{m}{2\pi\hbar^{2}}[1+\frac{K_{F}}{q}{\rm sign}\left(\alpha_{-}\right)\sqrt{\alpha_{-}^{2}-1}\dots-\frac{K_{F}}{q}{\rm sign}\left(\alpha_{+}\right)\sqrt{\alpha_{+}^{2}-1}]\label{eq:realfreqPolarization}
\end{equation}
where $\alpha_\pm=\frac{\omega^+ m}{q\hbar K_{F}}\pm\frac{q}{2K_{F}}$, $\omega^+=\omega+\imath 0^+$ and the square root has positive real part~\cite{Stern1967}. This is the analytical continuation, that is analytic in the upper half-plane, to the real frequencies of the Matsubara frequency form.

\begin{figure*}
    \includegraphics[width=\textwidth]{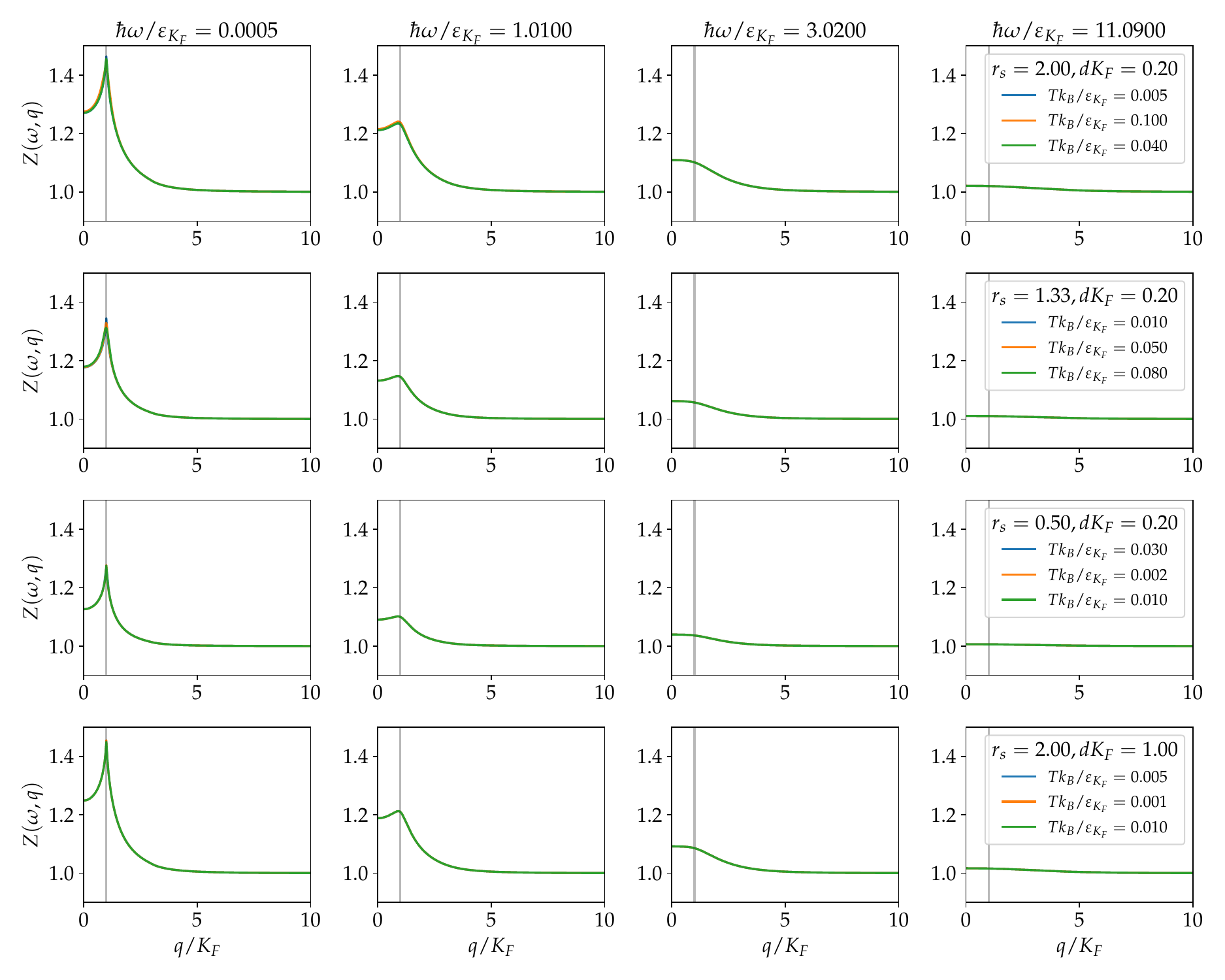}
    \caption{$Z_{\boldsymbol{q}}$ plotted as a function of $q$ and $\omega$ for different values of $r_s$, $T$ and $d$. Each column represents a cut along a fixed value of $\omega$. Each row corresponds to a fixed value of $r_s$ and $d$. The $T$ dependence of $Z_{\boldsymbol{q}}$ is weak enough that the results at different $T$ exactly overlap. The vertical line shows the Fermi wavevector $q/K_F=1$.\label{fig:Z}}
    \end{figure*}

\section{Solutions for normal self energy}

In this section we show further details of the numerical solutions of the linearized Eliashberg equations. Expanding Eq.~10 of the the main text to linear order in $W_{\boldsymbol{q}}$, the equations for the normal self-energy $\Sigma_{\boldsymbol{q}}$ (expressed in terms of its components $S_{\boldsymbol{q}}$ and $Z_{\boldsymbol{q}}$) are found to be independent of $W_{\boldsymbol{q}}$. This allows us to solve the equations for $Z_{\boldsymbol{q}}$ and $S_{\boldsymbol{q}}$ in Eq.~10 of the main text and use the solutions as inputs to the linear equation for $W_{\boldsymbol{q}}$.

\subsection*{Odd component of the normal self energy}

$Z_{\boldsymbol{q}}$, which determines the odd-part of $\Sigma_{\boldsymbol{q}}$, is nearly independent of the temperature at low temperatures and depends only weakly on $r_s$ and $d$. Figure~\ref{fig:Z} shows representative plots of $Z_{\boldsymbol{q}}$ as a function of $q$ for different values of $\omega$, $r_s$ and $d$. At large $q$ and $\omega$, $Z_{\boldsymbol{q}}$ retains its bare value of $1$. The corrections are, as expected, larger for larger $r_s$ and weakly increases with decreasing $d$. Deviation from one is maximum at the Fermi wavevector (i.e. $q/K_F\sim 1$). The results shown in Fig.~\ref{fig:Z} are qualitatively similar to those found in Ref.~\cite{Rietschel1983}.


\subsection*{Even component of the normal self energy}

\begin{figure*}
    \includegraphics[width=\textwidth]{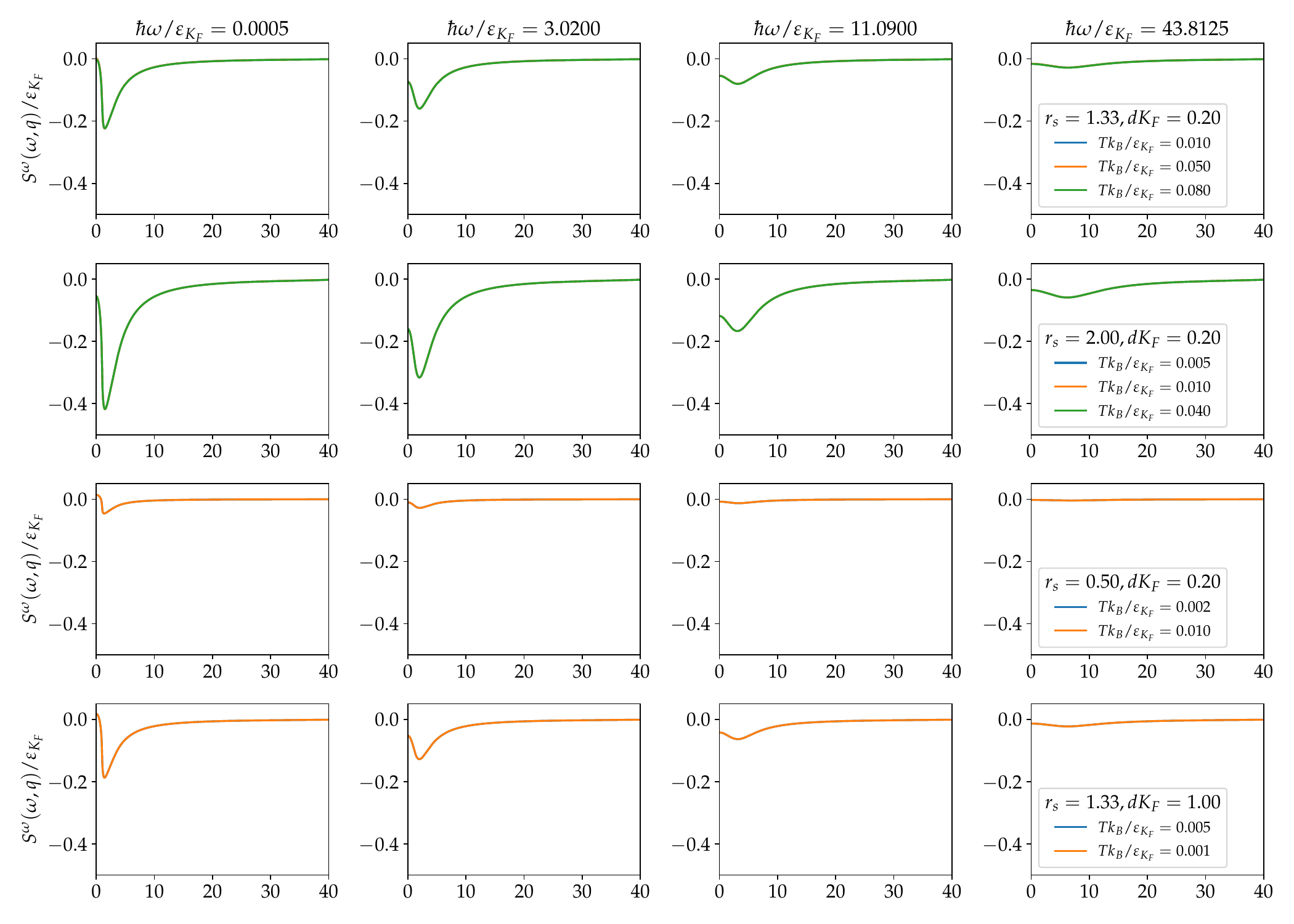}
    \caption{Frequency dependent even component of the normal self energy as a function of $q$ for different values of $\omega$ (different columns). Each row shows a fixed value of $r_s$ and $d$. Lines for different values of $T$ overlap indicating nearly temperature independent nature of the self-energy.\label{fig:SeOmega}}
\end{figure*}

The even component of the self energy is decomposed as $S_{\boldsymbol{q}} = S^{\omega}_{\boldsymbol{q}} + S^{0}_{{q}}$ where $S^\omega_{\boldsymbol{q}}$ is a frequency dependent part which we find to have a finite support in the $q-\omega$ plane and $S^0_q$ is a frequency independent part. The latter is written as a sum of two terms for numerical convenience
\begin{equation}
    S^{0}_{{q}} = S^{0,1}_{{q}} + S^{0,2}_{{q}}\nonumber
\end{equation}
where 
\begin{align}
    S_{{q}}^{0,1} &= \frac{1}{\hbar\beta}  \int_k V_{k-q} \sum_{\imath \omega_k} \left ( \frac{E_{\boldsymbol{k}}/\hbar}{\Lambda_{\boldsymbol{k}}} - \frac{E^0_{k}/\hbar}{\Lambda^0_{k}}\right)\label{eq:exchangeSelfEnergyNonSummable}\\
    S_{{q}}^{0,2} &= \frac{1}{\hbar \beta} \int_k V_{k-q} \left( -\frac{1}{2}+\sum_{\imath \omega_k} \frac{E^0_{k}/\hbar}{\Lambda^0_{k}} \right)\label{eq:exchangeSelfEnergySummable}
\end{align}
where $E_{k}^0=\varepsilon_k+S_{k,\imath \omega_k\to 0}-\mu$ and $\Lambda^0_{k}={[\omega_k Z_{k,\imath \omega_k\to 0}]}^2 + \frac{1}{\hbar^2} {{{(E_{k}^0)}^2}}$.
The integral over the angular part of $\vec{k}$ in Eq.~\ref{eq:exchangeSelfEnergyNonSummable} and Eq.~\ref{eq:exchangeSelfEnergySummable}, as well as the Matsubara summation in Eq.~\ref{eq:exchangeSelfEnergySummable} can be perfomed exactly allowing accurate and efficient estimation of $S^{0}$. 

Figure~\ref{fig:SeOmega} shows $S^\omega_{\boldsymbol{q}}$ as a function of $q$ for different values of $r_s$, $T$, $d$ and $\omega$. $S^\omega_{\boldsymbol{q}}$ vanishes at large $q$ and $\omega$. Thus $S^0_{\boldsymbol{q}}$ can be identified with the $\omega\to \infty$ limit of $S_{\boldsymbol{q}}$. 
$S^{\omega}_{\boldsymbol{q}}$ is nearly temperature independent. It is smaller (in magnitude) for weakly interacting systems (smaller $r_s$) and decreases in magnitude with increasing $d$.

\begin{figure*}
    \includegraphics[width=0.5\textwidth]{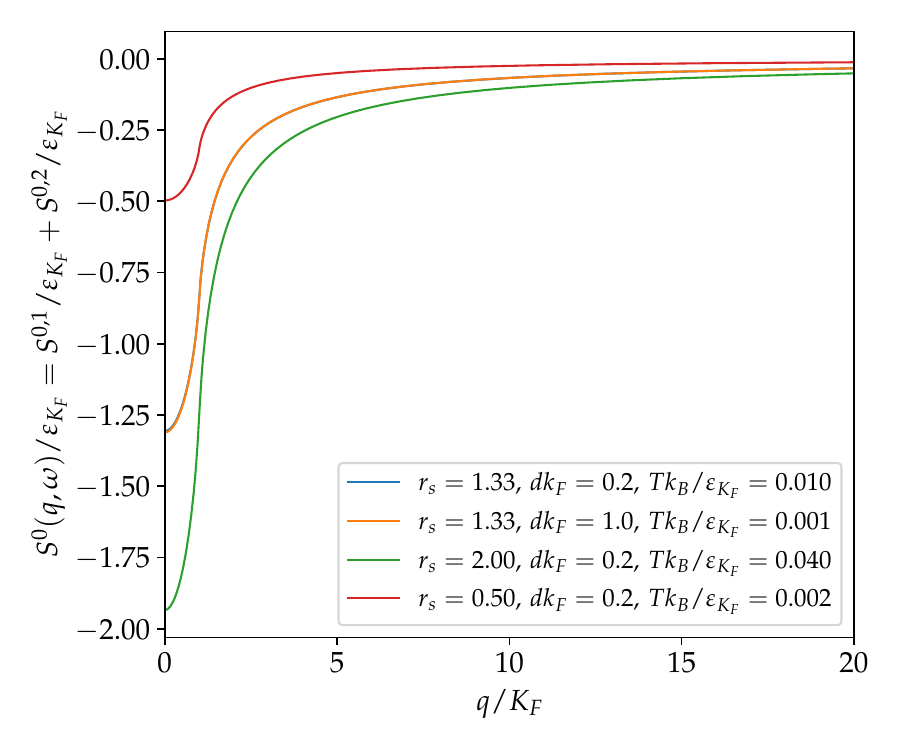}\includegraphics[width=0.5\textwidth]{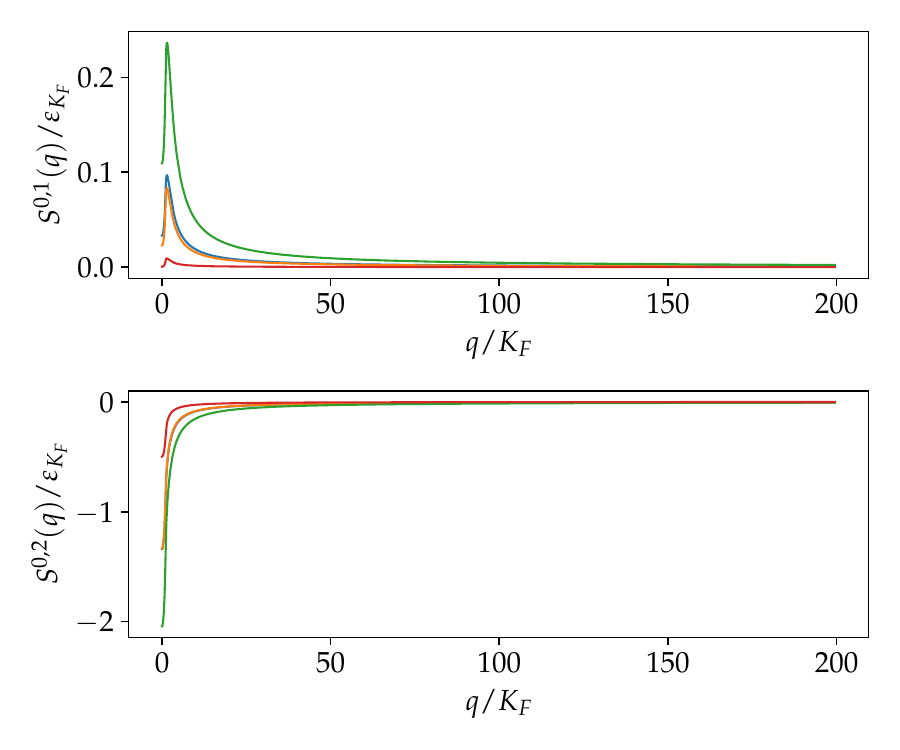}
    \caption{(left) Infinite frequency component $S^{0}$ of the even part of the normal self energy. (right) The top and bottom panels show the components $S^{0,1}$ and $S^{0,2}$ of $S^0$.~\label{fig:S012}}
\end{figure*}

Figure~\ref{fig:S012} shows the results for the infinite frequency component ($S^0$) of $S$ which was subtracted from $S$ to get $S^{\omega}$. As discussed above, $S^0$ is evaluated as a sum of two terms $S^{0,1}$ and $S^{0,2}$. $S^0$, $S^{0,1}$ and $S^{0,2}$ are shown as a function of $q$ for different values of $r_s$ and $d$ in the three panels. 
Temperature dependence is minimal and therefore not shown. Self-energies $S^{0,1}$ and $S^{0,2}$ are larger (in magnitude) when $r_s$ is larger (stronger interactions) and weakly increases with decreasing $d$. 

\begin{figure*}
    \includegraphics[width=0.5\textwidth]{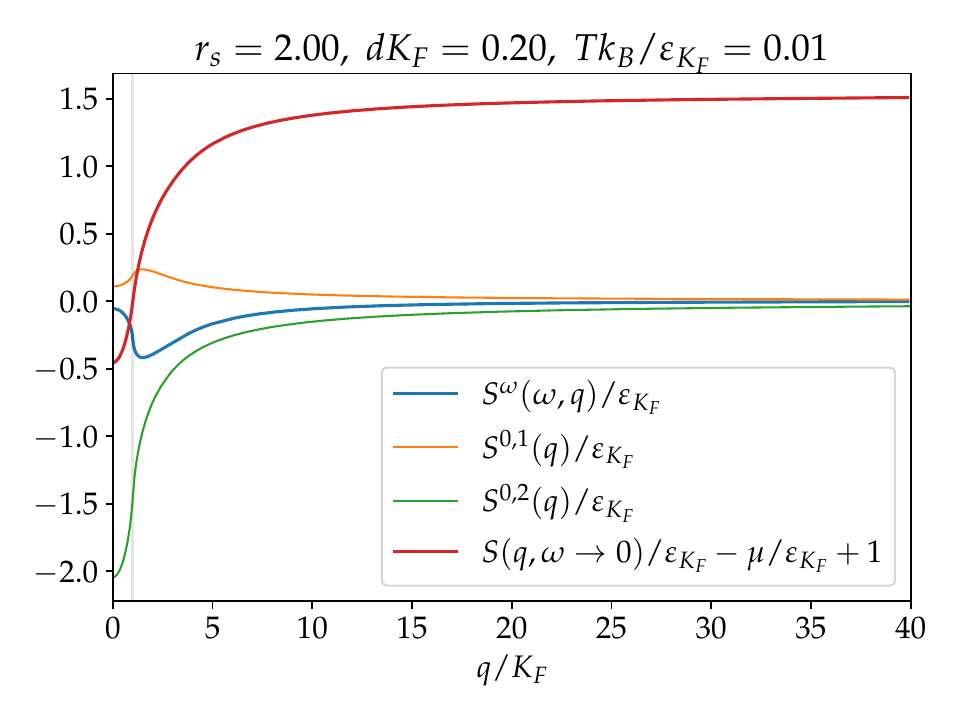}
    \caption{Different components of the even part of the normal self-energy as a function of $q$ for representative values of $r_s$ and $d$. Zero frequency component of the total $S_{\boldsymbol{q}}$ is shown in red.~\label{fig:totalSe}}
\end{figure*}

Figure~\ref{fig:totalSe} shows the different components of the even part of the self-energy as a function of $q$ for a representive case of $r_s=2$, $dK_F=0.2$ and $T/T_F=0.01$. The sum of all components and the chemical potential is also shown along the $\omega=0$ cut and agrees qualitatively with Ref.~\cite{Rietschel1983}. 

\section{Eigenfunctions of the linearized equation for \texorpdfstring{$W_{\boldsymbol{q}}$}{Lg}}

\begin{figure*}
    \includegraphics[width=\textwidth]{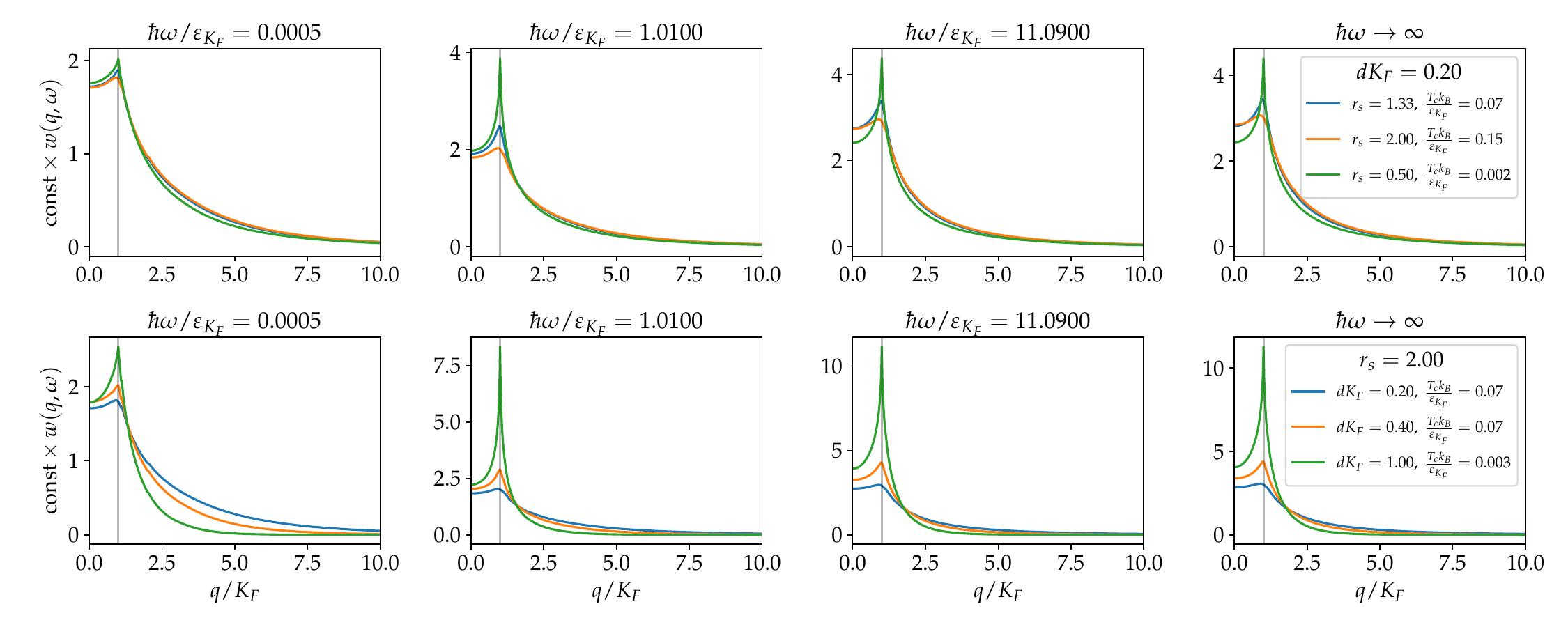}
    \caption{Eigenfunctions $w_{\boldsymbol{q}}$ of the leading eigenvalue of the linearized equation for $W$ at a temperature near $T_c$. ~\label{fig:W}}
\end{figure*}

In this section, we show the eigenfunctions of the linearized equation for $W_{\boldsymbol{q}}$ (Eq.~10 of the main text). This equation has the form $W=F[W]$ where $F$ is a $T$ dependent linear function of its argument $W$. When seen as an iterative equation, $W$ vanishes if the largest eigenvalue $\lambda$ of $F$ has a magnitude less than $1$. $|\lambda|$ is found to increase with decreasing $T$ and the transition temperature $T_c$ is identified by solving $|\lambda(T_c)|=1$.

Figure~\ref{fig:W} shows the eigenfunctions $w_{\boldsymbol{q}}$ corresponding to the largest (in magnitude) eigenvalue of $F$ at $T=T_c$. Note that the eigenfunctions are defined only upto a multiplicative constant and the magnitude of the functions shown is not a physically meaningful quantity. $w_{\boldsymbol{q}}$ has a peak near the Fermi wavevector for all $\omega$ which broadens for larger $r_s$ and smaller $d$. $w$ asymptotes to a nontrivial $q$ dependent function at $\omega$ which is shown in the last column of the Fig.~\ref{fig:W}.

\section{Fermi-surface averaged spectral function of the interaction\label{sec:U-spectral-function}}
\begin{figure}
\includegraphics[width=0.5\textwidth]{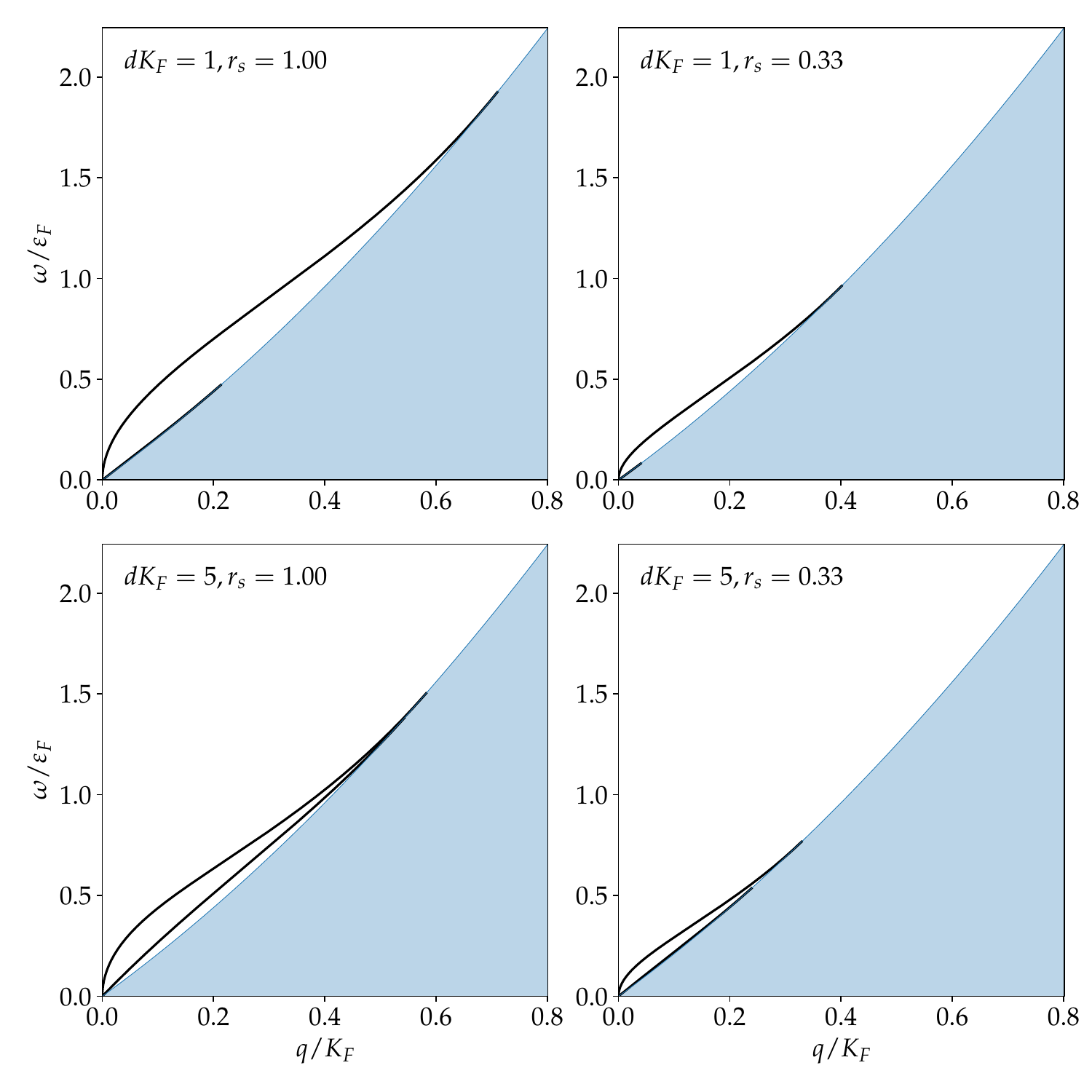}
\caption{Plasmon dispersion (black line) shown in the $q-\omega$ plane. The plasmon solutions exists for a range of small $q$ after which they dissipate away into the particle hole continuum (blue region). There are two plasmon modes which arise from the coupled pair of layers.\label{fig:plasmons}}
\end{figure}
The usefulness of the Eliashberg theory is limited to the cases where the frequencies contained in the attractive interaction are small relative to the Fermi energies. It is convenient to consider the real frequency form of the effective interaction given by 
\begin{equation}
    U^{\rm eff}_{\boldsymbol{q}} = \frac{1}{2}\sum_{\sigma=\pm 1} \sigma \frac{V_{\vec{q}}+\sigma U_{\vec{q}}}{1+(V_{\vec{q}}+\sigma U_{\vec{q}})\Pi^R_{\boldsymbol{q}}}.
\end{equation}
wherein $\Pi^R$ is given by Eq.~\ref{eq:realfreqPolarization}. Divergences of $U^{\rm eff}$ at the zeros of its denominator correspond to the plasmons. The plasmon dispersion for various choices of $d,r_s$ are shown in Fig.~\ref{fig:plasmons}. They exist only at low frequencies and this frequency regime decreases with decreasing $r_s$ and increasing $d$. The dominant contributions to $U^{\rm eff}$ are then expected to be only at low frequencies and these frequencies decrease with decreasing $r_s$ and increasing $d$. A qualitative understanding of the frequency content can be obtained by considering the spectral function of $U$ averaged over the fermi surface. We define this quantity as 
\begin{equation}
    \alpha^2 F = -\frac{1}{\pi}{\rm Im} \int_{q} U^{\rm eff}_{\boldsymbol{q}} \delta(|q|-K_F)
\end{equation}

\begin{figure}
\includegraphics[width=0.65\textwidth]{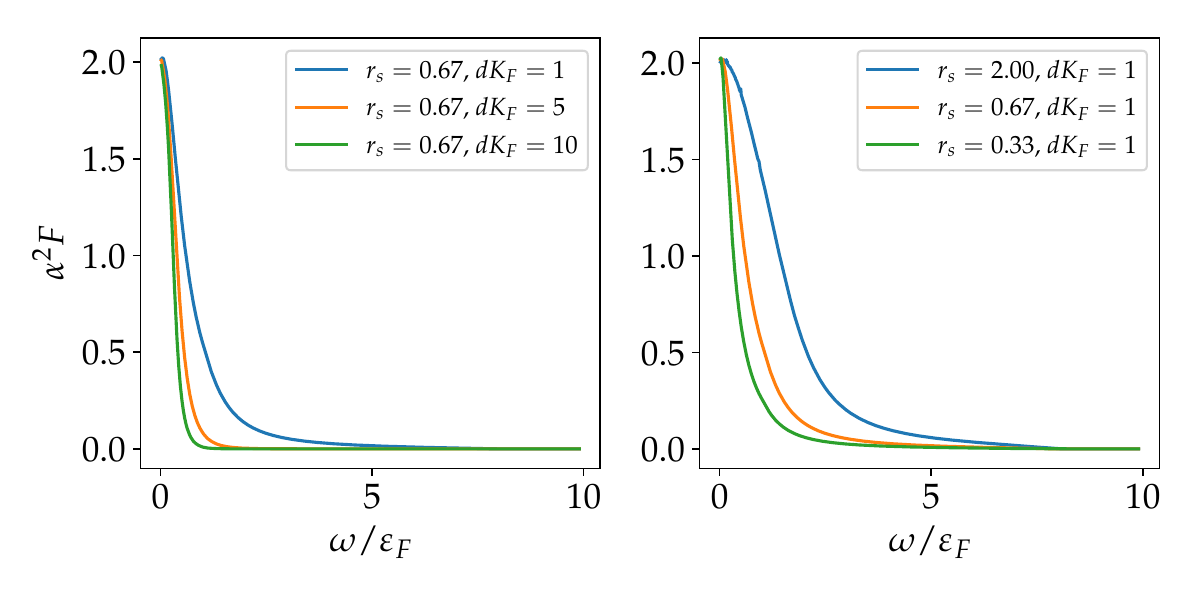}
\caption{Spectral function of $U^{\rm eff}$ for different values of $dK_F$ and $r_s$\label{fig:alpha2F}}
\end{figure}

Fig.~\ref{fig:alpha2F} shows this quantity as a function of $\omega/\varepsilon_F$ for different values of $r_s$ and $d$. The spectral function is substantial only at low frequencies and this frequency window decreases with decreasing $r_s$ and increasing $d$ consistent with the expectation that the plasmon peaks form the dominant contributions to the effective interaction. This suggests that Eliashberg theory should be reliable in the limit of small $r_s$ and large $d$.

\bibliography{exciton.bib}